\documentclass[twocolumn, showpacs, nofootinbib, aps, prd]{revtex4}

\usepackage{graphicx}
\usepackage{dcolumn}
\usepackage{bm}
\usepackage{hyperref}
\usepackage{color}

\begin{document}

\title{Determination of the pion distribution amplitude}

\author{Tao Huang$^1$}
\email{huangtao@ihep.ac.cn}
\author{Tao Zhong$^{1}$}
\email{zhongtao@ihep.ac.cn}
\author{Xing-Gang Wu$^{2}$}
\email{wuxg@cqu.edu.cn}

\address{$^1$ Institute of High Energy Physics and Theoretical Physics Center for Science Facilities, Chinese Academy of Sciences, Beijing 100049, P.R. China \\
$^{2}$ Department of Physics, Chongqing University, Chongqing 401331, P.R. China}

\date{\today}

\begin{abstract}

Right now, we have not enough knowledge to determine the hadron distribution amplitudes (DAs) which are universal physical quantities in the high energy processes involving hadron for applying pQCD to exclusive processes. Even for the simplest pion, one can't discriminate from different DA models. Inversely, one expects that processes involving pion can in principle provide strong constraints on the pion DA. For example, the pion-photon transition form factor (TFF) can get accurate information of the pion wave function or DA, due to the single pion in this process. However, the data from Belle and BABAR have a big difference on TFF in high $Q^2$ regions, at present, they are helpless for determining the pion DA. At the present paper, we think it is still possible to determine the pion DA as long as we perform a combined analysis of the most existing data of the processes involving pion such as $\pi \to \mu \bar{\nu}$, $\pi^0 \to \gamma \gamma$, $B\to \pi l \nu$, $D \to \pi l \nu$, and etc. Based on the revised light-cone harmonic oscillator model, a convenient DA model has been suggested, whose parameter $B$ which dominates its longitudinal behavior for $\phi_{\pi}(x,\mu^2)$ can be determined in a definite range by those processes. A light-cone sum rule analysis of the semi-leptonic processes $B \to \pi l \nu$ and $D \to \pi l \nu$ leads to a narrow region $B = [0.01,0.14]$, which indicate a slight deviation from the asymptotic DA. Then, one can predict the behavior of the pion-photon TFF in high $Q^2$ regions which can be tested in the future experiments. Following this way it provides the possibility that the pion DA will be determined by a global fit finally.

\end{abstract}


\maketitle

\section{Introduction}

In the perturbative QCD (pQCD) theory, the distribution amplitude (DA) provides the underlying links between the hadronic phenomena in QCD at the large distance (nonperturbative) and the small distance (perturbative). The pion DA is an important element for applying pQCD calculation to the exclusive processes in the high energy processes involving pion, and inversely, all of them can in principle provide strong constraints on the pion DA. The pion DA is usually arranged according to its different twist structures. There are processes in which the contributions from the higher twists are highly power suppressed at the short distance. For example, it has been found that the contribution to the pion-photon transition form factor (TFF) from higher helicity and higher twist structures is negligible~\cite{TFF2007,TFF1997}. Thus, those processes will provide good platforms to learn the properties of the leading-twist pion DA. It is well-known that the leading-twist DA has the definite asymptotic form, $\phi^{as}_\pi(x,\mu^2)|_{\mu^2\to\infty} = 6x(1-x)$, which is independent to its shape around some initial scale $\mu_0\sim {\cal O}(1 GeV)$. However, in practical calculation, it is important to know what is the right shape of the pion DA at low and moderate scales.

The pion leading-twist DA at any scale $\mu$ can be expanded in Gegenbauer series in the following form~\cite{QCD_evolution,rady}
\begin{eqnarray}
\phi_\pi(x, \mu^2) = 6x(1-x) \sum^\infty_{n=0} a_n(\mu^2) C^{3/2}_n(2x-1),
\label{DA_Gegenbauer}
\end{eqnarray}
where $C^{3/2}_n(2x-1)$ are Gegenbauer polynomials and the nonperturbative coefficients $a_n(\mu^2)$ are Gegenbauer moments. Due to the isospin-symmetry, only the even moments are non-zero. Usually the Gegenbauer series is convergent, one can adopt the first several terms to analyze the experimental data. If the shape of the pion DA at an initial scale $\mu_0$ is known, then
\begin{itemize}
\item by using the orthogonality relations for the Gegenbauer polynomials, the Gegenbauer moments $a_n(\mu^2_0)$ can be obtained via the equation,
    \begin{eqnarray}
     a_n(\mu^2_0) = \frac{\int^1_0 dx \; \phi_\pi(x,\mu^2_0) C^{3/2}_n(2x-1)} {\int^1_0 dx \; 6x(1-x)[C^{3/2}_n(2x-1)]^2}.  \label{Gegenbauer_moment}
    \end{eqnarray}
\item by using the QCD evolution equation~\cite{TFFas}, one can derive the pion DA at any other scale from $\phi_\pi(x,\mu^2_0)$.
\end{itemize}

The value of the Gegenbauer moments have been studied by using the non-perturbative approaches as the QCD sum rules~\cite{T2_SRs_earliest,T2_SRs_others} or the lattice
QCD~\cite{T2_Lattice}. However, at present, there is no definite conclusion on whether the pion DA $\phi_\pi(x,\mu^2_0)$ is asymptotic form~\cite{TFFas} or CZ-form~\cite{cz} or even flat-like~\cite{DA_flat}. It would be helpful to have a general pion DA model that can mimic all the DA behaviors suggested in the
literature. For this purpose, one can first construct a wavefunction (WF) model, since the pion DA is related to its WF $\Psi_\pi(x,\mathbf{k}_\bot)$ via the following relation,
\begin{eqnarray}
\phi_\pi(x,\mu^2_0) = \frac{2\sqrt{6}}{f_\pi} \int_{\left| \mathbf{k}_\bot \right|^2 \leq \mu^2_0} \frac{d^2\mathbf{k}_\bot}{16\pi^3} \Psi_\pi(x,\mathbf{k}_\bot) ,
\label{DA_WF}
\end{eqnarray}
where $f_{\pi}$ is the pion decay constant. It is noted that a proper way of constructing the pion WF/DA is also very important to derive a better end-point behavior at small $x$ and $k_\perp$ region for dealing with high energy processes within the $k_T$-factorization approach~\cite{kt}, and thus to provide a better estimation for the pion photon TFF, pion electromagnetic form factor, and etc.

The revised light-cone harmonic oscillator model for the pion leading-twist WF, and hence the model for the leading-twist DA, has been suggested in Refs.\cite{TFF2010,TFF2011,TFF2012}. It has been found that by a proper change of the pion DA parameters, one can conveniently simulate the shape of the DA from asymptotic-like to CZ-like. By comparing the theoretical estimations on the
pionic processes with the corresponding experimental data, those undetermined parameters of the DA model can be fixed or at least be greatly restricted. This is the purpose of the present paper.

More explicitly, we shall make a combined analysis of the pion DA by
using the pion decay channels $\pi^0 \to \gamma \gamma$ and $\pi \to
\mu \bar{\nu}$, the pion-photon TFF $F_{\pi\gamma}(Q^2)$, the
semi-leptonic decays $B\to \pi l \nu$ and $D \to \pi l \nu$, and the
exclusive process $B^0 \to \pi^0 \pi^0$. For example, the pion-photon TFF
$F_{\pi\gamma}(Q^2)$ that relates pion with two photons provides the
simplest example for the perturbative application to exclusive
processes. In the lower energy region the data on the pion-photon
TFF measured by CELLO, CLEO, BABAR and Belle are consistent with
each other~\cite{TFF_CELLO,TFF_CLEO,TFF_BABAR,TFF_BELLE}, so these
data can be adopted for constraining the WF parameters. Based on the
present DA model, the model parameter $B$ for $\phi_{\pi}(x,\mu^2)$
can be determined, then one can predict the behavior of the
pion-photon TFF in high $Q^2$ regions which can be tested in the
future experiments.

The remaining parts of the paper is organized as follows. In Sec.II,
we give a brief review on the pion leading-twist WF/DA, properties
of DA have also been presented there. In Sec.III, we show how DA
parameters can be constrained, and present a detailed derivation of the
parameter $B$ by using the $B/D \to \pi$ transition form factors
within the light-cone sum rule (LCSR). A discussion on the pion-photon TFF and $B^0 \to
\pi^0 \pi^0$ process is presented in Sec.V. The
final section is reserved for a summary.

\section{A Brief Review on the Pion leading-twist WF/DA}

One useful way for modeling the hadronic valence WF is to use the
approximate bound state solution of a hadron in terms of the quark
model as the starting point. The Brodsky-Huang-Lepage (BHL)
prescription~\cite{BHL} of the hadronic WF is rightly
obtained in this way by connecting the equal-time WF in
the rest frame and the WF in the infinite momentum frame.
Based on this prescription, the revised light-cone harmonic
oscillator model of the pion leading-twist WF has suggested in
Refs.\cite{TFF2010,TFF2011}, which shows
\begin{eqnarray}
\Psi_\pi(x,\textbf{k}_\bot) = \sum_{\lambda_1 \lambda_2} \chi^{\lambda_1 \lambda_2}(x,\textbf{k}_\bot) \Psi^R_\pi(x,\textbf{k}_\bot),
\label{WF_full}
\end{eqnarray}
where $\chi^{\lambda_1 \lambda_2}(x,\textbf{k}_\bot)$ stands for the spin-space WF, $\lambda_1$ and $\lambda_2$ being the helicity states of the two constitute quarks in pion. The $\chi^{\lambda_1 \lambda_2}(x,\textbf{k}_\bot)$ comes from the Wigner-Melosh rotation whose explicit form can be found in Refs.\cite{WF94,WF_spin}. $\Psi^R_\pi(x,\textbf{k}_\bot)$ indicates the spatial WF, which can be divided into a $\textbf{k}_\bot$-dependent part and a $x$-dependent part. For the $\textbf{k}_\bot$-dependent part, Brodsky-Huang-Lepage suggests that there is possible connection between the rest frame WF $\Psi_{c.m}(\textbf{q})$ and the light-cone WF $\Psi_{LC}(x,\textbf{k}_\bot)$~\cite{BHL}:
\begin{eqnarray}
\Psi_{c.m.}(\textbf{q}^2) \longleftrightarrow \Psi_{LC} \left[ \frac{\textbf{k}_\bot^2 + m^2_q}{4x(1-x)} - m^2_q \right],
\label{WF_connection}
\end{eqnarray}
where $m_q$ stands for the mass of the constitute quarks. From an approximate bound-state solution in the quark models for pion, the WF of the harmonic oscillator model in the rest frame can be obtained~\cite{WF_restframe}. Thus, for the $\textbf{k}_\bot$-dependent part of spatial WF $\Psi^R_\pi(x,\textbf{k}_\bot)$, we have:
\begin{eqnarray}
\Psi^R_\pi(x,\textbf{k}_\bot) \propto \exp \left[ -\frac{\textbf{k}^2_\bot + m_q^2}{8\beta^2 x(1-x)} \right]. \label{WF_LC}
\end{eqnarray}
For the $x$-dependent part of $\Psi^R_\pi(x,\textbf{k}_\bot)$, we take $\varphi_\pi(x) = [1 + B \times C^{3/2}_2(2x-1)]$, which dominates the longitudinal distribution broadness of the WF and can be expanded in the Gegenbauer polynomials. Here we only keep the first two terms in $\varphi_\pi(x)$, in which the parameter $B\sim a_2$ can be regarded as an effective parameter to determine the broadness of the longitudinal part of the WF.

As a combination, the explicit form of the spatial WF can be obtained:
\begin{eqnarray}
\Psi^R_\pi(x,\textbf{k}_\bot) = A \varphi_\pi(x) \exp \left[ -\frac{\textbf{k}^2_\bot + m_q^2}{8\beta^2 x(1-x)} \right], \label{WF_spatial}
\end{eqnarray}
where $A$ is the normalization constant. After integration over the transverse momentum dependence, one can obtain the pion DA with the help of Eq.(\ref{DA_WF}),
\begin{widetext}
\begin{eqnarray}
\phi_\pi(x,\mu^2_0) = \frac{\sqrt{3} A m_q \beta}{2\pi^{3/2}f_\pi} \sqrt{x(1-x)} \varphi_\pi(x) \times \left\{ \textrm{Erf}\left[ \sqrt{\frac{m_q^2 + \mu_0^2}{8\beta^2 x(1-x)}} \right] - \textrm{Erf}\left[ \sqrt{\frac{m_q^2}{8\beta^2 x(1-x)}} \right] \right\}, \label{DA_model}
\end{eqnarray}
\end{widetext}
where $\textrm{Erf}(x) = \frac{2}{\sqrt{\pi}} \int^x_0 e^{-t^2} dt$.

Except for the constitute quark mass $m_q$, which can be taken as the conventional value about $0.30 \textrm{GeV}$, there are three undetermined parameters, $A$, $\beta$ and $B$, in the above model. Two important constraints have been found in Ref.\cite{BHL} to constrain those parameters: (1) the process $\pi \to \mu \bar{\nu}$ provides the WF normalization condition
\begin{eqnarray}
\int^1_0 dx \int \frac{d^2 \textbf{k}_\bot}{16\pi^3} \Psi_\pi(x,\textbf{k}_\bot) = \frac{f_\pi}{2\sqrt{6}};
\label{DA_constraint1}
\end{eqnarray}
(2) the sum rule derived from $\pi^0 \to \gamma\gamma$ decay amplitude implies,
\begin{eqnarray}
\int^1_0 dx \Psi_\pi(x, \textbf{k}_\bot = \textbf{0}) = \frac{\sqrt{6}}{f_\pi}.
\label{DA_constraint2}
\end{eqnarray}

\begin{figure}
\centering
\includegraphics[width=0.45\textwidth]{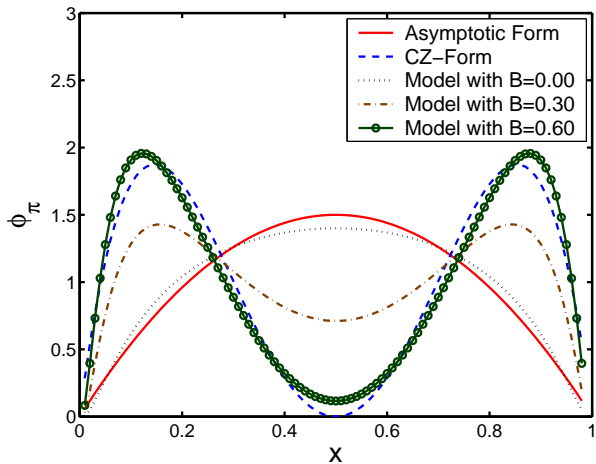}
\caption{The pion DA model $\phi_\pi(x,\mu_0^2)$ defined in Eq.(\ref{DA_model}) versus the parameter $B$~\cite{TFF2010}. By varying $B$ from $0.00$ to $0.60$, $\phi_\pi(x,\mu_0^2)$ changes from asymptotic-like to CZ-like. } \label{phi}
\end{figure}

In addition to these two basic constraints, one needs other processes involving pion to further constrain the parameters, especially to determine the value of the parameter $B$. We put the DAs for $B=0.00$, $0.30$ and $0.60$ in Fig.(\ref{phi}), where as a comparison, the asymptotic DA and CZ-DA have also been present. When the value of $B$ changes from $0.00$ to $0.60$, together with the constraints (\ref{DA_constraint1}) and (\ref{DA_constraint2}), the pion DA model can mimic the DA shapes from asymptotic-like to CZ-like:
\begin{itemize}
\item The second moments $a_{2}$ varies from $0.03$ to $0.68$;
\item The first inverse moments of the pion DA at energy scale $\mu_0$, $\int_0^1 \left[\phi_\pi(x,\mu_0^2)/x \right] dx$, varies from $3.0$ to $5.0$. \\
\end{itemize}
Thus, if we have precise measurements for certain processes, then by comparing the theoretical estimations derived under the DA model (\ref{DA_model}), one can conveniently fix the pion DA behavior.

\begin{table}[htb]
\caption{Typical pion WF parameters with $m_q=0.30$ GeV and $\mu_0=1$ GeV. The second and fourth Gegenbauer moments are also presented.}
\begin{center}
\begin{tabular}{c c c c c c}
\hline ~~~$B$~~~ & ~$A ({\rm GeV}^{-1})$~& ~$\beta ({\rm GeV})$~ & ~$a_2(\mu_0^2)$~ & ~$a_4(\mu_0^2)$~  \\
\hline
~$-0.60$~ & ~$36.03$~& ~$0.456$~ & ~$-0.523$~ & ~$0.051$~  \\
\hline
~$-0.30$~ & ~$30.43$~& ~$0.514$~ & ~$-0.279$~ & ~$0.000$~  \\
\hline
~$0.00$~ & ~$24.80$~& ~$0.589$~ & ~$0.028$~ & ~$-0.027$~  \\
\hline
~$0.30$~ & ~$20.05$~& ~$0.672$~ & ~$0.364$~ & ~$-0.017$~  \\
\hline
~$0.60$~ & ~$16.46$~& ~$0.749$~ & ~$0.681$~ & ~$0.022$~   \\
\hline
\end{tabular}
\label{tab1}
\end{center}
\end{table}

We put the WF parameters for several typical $B$ in Table \ref{tab1}, where the region of the parameter $B$ is broadened to be $[-0.60,0.60]$. The value of $B$ is close to the second Gegenbauer moment, $B\sim a_{2}$, and because of the fact that the longitudinal distribution is dominated by the second Gegenbauer moment, cf.Refs.\cite{TFFas,cz,BPI_TFF05,BPI_TFF08,BPI_TFF11,PP_TFF}, thus the parameter $B$ dominantly determines the broadness of the longitudinal part of the wave function.

The parameters listed in Table \ref{tab1} are for $\mu_0 = 1$ GeV. They can be run to any other scales by applying the evolution equation, i.e. to order ${\cal O}(\alpha_s)$, we have~\cite{TFFas}
\begin{widetext}
\begin{equation}
x_1 x_2 \frac{\partial \tilde{\phi}_{\pi}(x_i,\mu^2)}{\partial \ln\mu^2}=C_F \frac{\alpha_s(\mu^2)}{4\pi} \left\{\int_0^1 [dy] V(x_i,y_i) \tilde{\phi}_{\pi}(y_i,\mu^2) - x_1 x_2 \tilde{\phi}_{\pi}(x_i,\mu^2) \right\}, \label{eq:evolution}
\end{equation}
where $[dy]= dy_1 dy_2 \delta(1- y_1 -y_2)$, $\phi_{\pi}(x_i,\mu^2) = x_1 x_2 \tilde{\phi}_{\pi}(x_i,\mu^2)$ and
\begin{eqnarray}
V(x_i,y_i) &=& 2 \left[x_1 y_2 \theta(y_1 - x_1) \left(\delta_{h_1\bar{h_2}}
+ \frac{\Delta}{(y_1 - x_1)} \right) + (1 \leftrightarrow 2) \right]. \nonumber \end{eqnarray}
\end{widetext}
The $\theta$ function is the usual step function, the color factor $C_F=4/3$, $\delta_{h_1\bar{h_2}}=1$ when the $q$ and $\bar{q}$ helicities are opposite, and $\Delta\tilde{\phi}_{\pi}(y_i,\mu^2)=
\tilde{\phi}_{\pi}(y_i,\mu^2) - \tilde{\phi}_{\pi}(x_i,\mu^2)$.

Practically, the above evolution (\ref{eq:evolution}) can be solved by using the DA Gegenbauer expansion (\ref{DA_Gegenbauer}), which transforms the DA scale dependence to the determination of the scale dependent of the Gegenbauer moments~\cite{QCD_evolution,rady}. More explicitly, the explicit expression for $a_n(\mu^2)$ to leading-logarithmic (LL) accuracy can be written as~\cite{PB2005}:
\begin{eqnarray}
a_n(\mu^2) = a_n(\mu^2_0) \left( \frac{\alpha_s(\mu^2)}{\alpha_s(\mu_0^2)} \right)^{\gamma_n/\beta_0},
\label{an_evolution}
\end{eqnarray}
where the anomalous dimensions
\begin{eqnarray}
\gamma_n = C_F \left( 1- \frac{2}{(n+1)(n+2)} + 4\sum^{n+1}_{m=2} \frac{1}{m} \right)
\label{anomalous_dimensions}
\end{eqnarray}
with $\beta_0 = (11N_c - 2N_f)/3$. Usually, one truncates the Gegenbauer expansion with the first several terms ($n=0,2,4,6$ respectively) to derive the DA behavior at the high energy scales.

\begin{table*}[htb]
\caption{Typical pion WF parameters for $m_q=0.30$ GeV at several typical energy scales, $\mu =1, 1.5, 3 {\rm GeV}$, respectively.}
\begin{tabular}{ c c c | c c c | c c c}
\hline\hline  \begin{tabular}{c} \quad \\ ~$A ({\rm GeV}^{-1})$~ \end{tabular} & \begin{tabular}{c} $\mu=1{\rm GeV}$ \\ $B$ \end{tabular} & \begin{tabular}{c} \quad \\ $\beta ({\rm GeV})$
 \end{tabular} & \begin{tabular}{c} \quad \\ ~$A ({\rm GeV}^{-1})$~ \end{tabular} & \begin{tabular}{c} $\mu=1.5{\rm GeV}$ \\ $B$ \end{tabular} & \begin{tabular}{c} \quad \\ $\beta ({\rm GeV})$
 \end{tabular} & \begin{tabular}{c} \quad \\ ~$A ({\rm GeV}^{-1})$~ \end{tabular} & \begin{tabular}{c} $\mu=3{\rm GeV}$ \\ $B$ \end{tabular} & \begin{tabular}{c} \quad \\ $\beta ({\rm GeV})$
 \end{tabular} \\
\hline
~$24.63$~& ~$0.01$~ & ~$0.592$~ & ~$24.99$~& ~$0.037$~ & ~$0.560$~ & ~$25.11$~& ~$0.033$~ & ~$0.556$~ \\
~$23.93$~& ~$0.05$~ & ~$0.603$~ & ~$24.40$~& ~$0.073$~ & ~$0.567$~ & ~$24.63$~& ~$0.062$~ & ~$0.562$~ \\
~$23.09$~& ~$0.10$~ & ~$0.617$~ & ~$23.67$~& ~$0.118$~ & ~$0.577$~ & ~$24.05$~& ~$0.099$~ & ~$0.570$~ \\
~$22.44$~& ~$0.14$~ & ~$0.628$~ & ~$23.11$~& ~$0.154$~ & ~$0.585$~ & ~$23.59$~& ~$0.128$~ & ~$0.576$~ \\
~$22.28$~& ~$0.15$~ & ~$0.631$~ & ~$22.97$~& ~$0.163$~ & ~$0.587$~ & ~$23.48$~& ~$0.135$~ & ~$0.578$~ \\
~$21.50$~& ~$0.20$~ & ~$0.645$~ & ~$22.30$~& ~$0.208$~ & ~$0.597$~ & ~$22.93$~& ~$0.171$~ & ~$0.585$~ \\
~$20.76$~& ~$0.25$~ & ~$0.658$~ & ~$21.65$~& ~$0.252$~ & ~$0.607$~ & ~$22.39$~& ~$0.207$~ & ~$0.593$~ \\
~$20.05$~& ~$0.30$~ & ~$0.672$~ & ~$21.03$~& ~$0.296$~ & ~$0.617$~ & ~$21.88$~& ~$0.242$~ & ~$0.601$~ \\
~$19.37$~& ~$0.35$~ & ~$0.686$~ & ~$20.43$~& ~$0.340$~ & ~$0.626$~ & ~$21.38$~& ~$0.277$~ & ~$0.608$~ \\
~$18.72$~& ~$0.40$~ & ~$0.699$~ & ~$19.87$~& ~$0.383$~ & ~$0.636$~ & ~$20.90$~& ~$0.311$~ & ~$0.616$~ \\
~$18.47$~& ~$0.42$~ & ~$0.704$~ & ~$19.65$~& ~$0.400$~ & ~$0.640$~ & ~$20.72$~& ~$0.325$~ & ~$0.618$~ \\
\hline\hline
\end{tabular}
\label{tab2}
\end{table*}

In this paper we solve the evolution equation (\ref{eq:evolution}) strictly to get the DA's behavior at the higher energy scale. It is noted that if the Gegenbauer expansion converges quickly, these two evolution methods (\ref{eq:evolution}) and (\ref{an_evolution}) are equivalent to each other. The solution of the evolution equation (\ref{eq:evolution}) can be done numerically. Here we suggest an
equivalent but simpler and more effective way to get the DA after evolution, i.e. we transform the whole scale dependence of $\phi_\pi(x,\mu^2_0)$ into the scale dependence of the undetermined parameters $A$, $B$ and $\beta$. The valence quark mass $m_q$ is scale independent and we keep it to be $0.30$ GeV. Its main idea is
to take the second Gegenbauer moment $a_2(\mu^2)$ as a ligament between the DA and the DA parameters. Firstly, from the initial DA $\phi_\pi(x,\mu^2_0)$ with known $A$, $B$ and $\beta$ at the initial $\mu_0$, we derive its second Gegenbauer moment $a_2(\mu^2_0)$ via Eq.(\ref{Gegenbauer_moment}), and get its value at any scale $\mu$ by using the evolution equation (\ref{an_evolution}). Secondly, we use the value of $a_2(\mu^2)$ together with the two constraints (\ref{DA_constraint1}) and (\ref{DA_constraint2}) to determine the values of $A$, $\beta$ and $B$ at the scale $\mu$. We put the parameters $A$, $B$ and $\beta$ at three typical scales $\mu =1$, $1.5$ and $3$ GeV in Table \ref{tab2}. From the table, one observes that the value of $A$ increases and the value of $\beta$ decreases with the increment of the scale.

\section{Determination of DA from $B/D \to \pi$ transition form factors}

The semi-leptonic $B$-meson decay $B\to\pi l \nu$ is usually used to extract the CKM matrix element $|V_{ub}|$, whose differential cross section for massless leptons can be written as
\begin{widetext}
\begin{eqnarray}
\frac{d\Gamma}{dq^2}(B\to\pi l \nu) && = \frac{C_F^2 |V_{ub}|^2}{192\pi^3 m_B^3} \left[ (q^2 + m_B^2 - m_\pi^2)^2
- 4m_B^2 m_\pi^2 \right]^{3/2} |f^{B\to\pi}_+(q^2)|^2,
\label{extract_vub}
\end{eqnarray}
\end{widetext}
where the momentum transfer $q = p_B - p_\pi$. The TFF $f^{B\to\pi}_+$ is the key factor of the process, which has been deeply investigated by using several approaches, such as the pQCD approach~\cite{BPI_TFF_PQCD,BPI_TFF_whole_region}, the QCD LCSR approach~\cite{PB2005,BPI_TFF05,BPI_TFF08,BPI_TFF11,BPI_TFF_T234,BPI_TFF_T2_LO,BPI_TFF_T2_NLO,BPI_TFF_T2,BPIT2_09,BDPI_TFF_LI,BPI_TFF_T3} and the lattice QCD approach~\cite{BPI_TFF_Lattice,Vub}. Different approaches are applicable for different energy regions. Among them, the QCD LCSR is reliable for the intermediate energy region, which can be extended to the whole physical region with proper extrapolation. So this approach is usually adopted for a detailed analysis in comparison with the experimental data.

Under LCSR, the expression for $f^{B\to\pi}_{+}$ depends on how one chooses the correlator~\cite{DA_fbpi}: different choice of the currents in the correlation function shall result in different expressions, in which, the pionic different twist structures provide different contributions. Here we adopt the chiral correlator suggested in Ref.\cite{BPI_TFF_T2_LO} to do our discussion, in which the leading-twist DA's contribution have been amplified and it provides us a better chance to know the detail of the leading-twist DA in comparison with data. By using the chiral correlator, up to twist-4, the form factor $f^{B\to\pi}_+(0)$ at the large recoil region can be obtained~\cite{BDPI_TFF_LI},
\begin{widetext}
\begin{eqnarray}
f_B f^{B\to\pi}_+(0) e^{-\frac{m_B^2}{M^2}} &=& -\frac{m_b^2 f_\pi}{2\pi m_B^2} \int^{s^B_0}_{m_b^2} ds e^{-\frac{s}{M^2}} \frac{1}{s} \int^{s/m^2_b}_0 d\eta \;{\rm Im} T\left( \frac{m_b^2}{s}\eta, \frac{s}{m_b^2}, \mu \right)  \phi_\pi \left( \frac{m_b^2}{s}\eta, \mu \right) \nonumber\\
&&  + \frac{f_\pi}{m_B^2} \int^1_{u_0} du e^{-\frac{m_b^2}{uM^2}} \left[ -\frac{u}{4} \frac{d^2 \phi_{4\pi}(u)}{du^2} + u\psi_{4\pi}(u) + \int^u_0 dv \psi_{4\pi}(v) - \frac{d}{du} I_{4\pi}(u) \right],
\label{fbpi}
\end{eqnarray}
\end{widetext}
where $f_B$, $m_B$, $M^2$, $m_b$ and $s^B_0$ indicate the $B$ meson decay constant, the $B$ meson mass, the Borel parameter, the $b$ quark mass and the effective threshold parameter, respectively. The parameter $u_0 = m^2_b / s_0^B$. The functions $\phi_{4\pi}$ and $\psi_{4\pi}$ are pion two-particle twist-4 DAs. $I_{4\pi}$ is a combination of pion three-particle twist-4 DAs. The hard scattering amplitude ${\rm Im}T\left( \eta m_b^2/s, s/m_b^2, \mu \right)$ involves the LO and NLO parts. The scale of the process $\mu = \sqrt{m_B^2 - m_b^2} \simeq 3$ GeV.

Furthermore, the semileptonic $D$-meson decay $D \to \pi l \nu$ can also be used to extract the CKM matrix element $V_{cd}$ if we know the $D\to\pi$ TFF $f^{D\to\pi}_+$ well. The TFF $f^{D\to\pi}_+$ has been studied in Refs.\cite{BDPI_TFF_LI,DPI_TFF_LCSR,DPI_TFF_Lattice}. Replacing all the $B$ meson parameters in (\ref{fbpi}) by those of $D$ meson, we can obtain the LCSR expression for $f^{D\to\pi}_+(0)$. For example, the scale for $f^{B\to\pi}_+$ now equals to $\mu = \sqrt{m_D^2 - m_c^2} \simeq 1.5$ GeV.

Using the formula (\ref{fbpi}), we obtain that the contributions from pion twist-4 DAs terms are less than $1\%$ for $f^{B \to\pi}_+(0)$ and less than $5\%$ for $f^{D \to\pi}_+(0)$. Thus this provides a good platform to study the properties of the pion leading-twist DA. In Ref.\cite{DA_fbpi}, the authors have made use of this platform to determine the DA parameter $B$ with experimental data of $f^{B\to\pi}_+ |V_{ub}|$ by taking the input parameters as same as in Ref.\cite{BDPI_TFF_LI}. At the present section, we update the analysis there by using the input parameters to be those given by the Particle Data Group~\cite{PDG}, and simultaneously we make
use of $f_{D^+}f^{D\to\pi}_+(0)$ as a further constrain to determine the pion DA parameters.

The input parameters are listed in the following. The $\overline{MS}$-running $b$ and $c$ masses, the $B^+$ and $D^+$ meson masses are~\cite{PDG}: $\bar{m}_b(\bar{m}_b) = 4.18 \pm 0.03$ GeV, $\bar{m}_c(\bar{m}_c) = 1.275 \pm 0.025$ GeV, $m_{B^+} = 5279.25 \pm 0.17$ MeV and $m_{D^+} = 1869.62 \pm 0.15$ MeV. The $B^+$ meson decay constant $f_{B^+} = 214^{-5}_{+7} {\rm MeV}$~\cite{BPI_TFF08}. Because there is large discrepancy for the estimation of the $D^+$ meson decay constant $f_{D^+}$~\cite{FD_FC,FD_Lattice,FD_LF,FD_PQL,FD_SR}, instead of using $f^{D\to\pi}_+(0)$ as a criteria, we adopt the combined value of $f_{D^+} f^{D\to\pi}_+(0)$ to constrain the pion DA. The pion decay constant $f_{\pi^0}$ is set to be $f_{\pi^+}$~\cite{PI_decayconstant}, which is $130.41 \pm 0.03 \pm 0.20$ MeV~\cite{PDG}. As for the effective threshold and Borel variables, we take them to be same as those of Ref.\cite{BDPI_TFF_LI}.

Experimentally, from the processes $B^+/D^+ \to \pi^0 l^+ \nu_l$, it has been shown that the multiplication of the form factor and the corresponding CKM matrix element by the BABAR~\cite{FVub} and CLEO~\cite{FVcd_CLEO} collaborations are,
\begin{eqnarray}
f^{B\to\pi}_+(0) |V_{ub}| = (9.4 \pm 0.3 \pm 0.3) \times 10^{-4}  \label{fvub}
\end{eqnarray}
and
\begin{eqnarray}
f^{D\to\pi}_+(0) |V_{cd}| = 0.146 \pm 0.007 \pm 0.002.   \label{fvcd}
\end{eqnarray}

\begin{figure}[ht]
\centering
\includegraphics[width=0.45\textwidth]{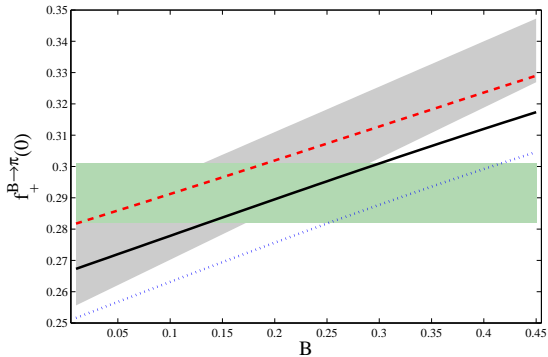}
\caption{The value of $f^{B\to\pi}_+(0)$ versus the DA parameter $B$. The solid, dashed and doted lines stand for central, upper and lower values obtained from the LCSR (\ref{fbpi}) with the leading-twist DA model (\ref{DA_model}). The lighter shaded band indicates the experimental band (\ref{exfbpi}). The thicker shaded band is the result of Ref.\cite{DA_fbpi}.} \label{fig_FBPI}
\end{figure}

From a simultaneous fit to the experimental partial rates and lattice points on the exclusive process $B \to \pi l \nu$ versus $q^2$, the CKM matrix element $|V_{ub}|$ is derived as $(3.23 \pm 0.31) \times 10^{-3}$~\cite{Vub}. As a combination, we can obtain the experimental value for $f^{B \to\pi}_+(0)$:
\begin{eqnarray}
f^{B\to\pi}_+(0) = 0.291^{+0.010}_{-0.009}.   \label{exfbpi}
\end{eqnarray}
Comparing this value with the estimated one from the LCSR (\ref{fbpi}), as indicated by Fig.(\ref{fig_FBPI}), we obtain the first reasonable region for the parameter $B$:
\begin{eqnarray}
B_{(B \to \pi l \nu)} = [0.01,0.42] ,  \label{B_Bpi}
\end{eqnarray}
where all the input parameters are varied within their reasonable regions listed above. Our present value for $B$ is different from that of Ref.\cite{BDPI_TFF_LI}, which is because we have adopted a different $\overline{MS}$ $b$-quark mass. Fig.(\ref{fig_FBPI}) gives the value of $f^{B\to\pi}_+(0)$ versus the parameter $B$. Where the lighter shaded band indicates the experimental value
(\ref{exfbpi}), the solid, dashed and doted lines stand for the central, upper and lower ones calculated by the LCSR (\ref{fbpi}), and the thicker shaded band is the result of Ref.\cite{DA_fbpi}.

\begin{figure}[ht]
\centering
\includegraphics[width=0.45\textwidth]{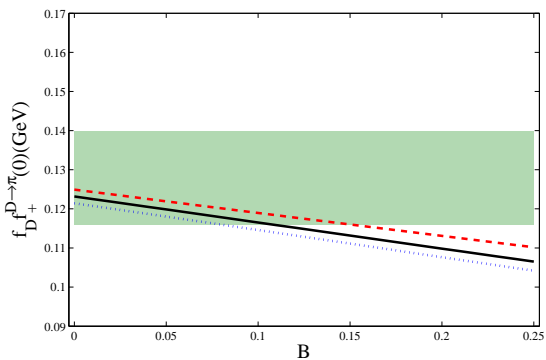}
\caption{The value of $f_{D^+}f^{D\to\pi}_+(0)$ versus the parameter $B$. The solid, dashed and doted lines stand for central, upper and lower values calculated by the LCSR (\ref{fbpi}) (with parameter changes for the $D$-meson case). The shaded band indicates the experimental value (\ref{exffdpi}) of $f_{D^+}f^{D\to\pi}_+(0)$.} \label{fig_FFDPI}
\end{figure}

For the $D$ meson case, whose lifetime is $1040 \pm 7 {\rm
fs}$~\cite{D_lifetime}, we can adopt the measurement of
$\mathcal{B}(D^+ \to \mu^+ \nu)$. Then using the formulae
\begin{displaymath}
\Gamma(D^+ \to l^+ \nu) = \frac{C^2_F}{8\pi} f^2_{D^+} m_l^2 m_{D^+} \left( 1 - \frac{m_l^2}{m_{D^+}^2} \right)^2 |V_{cd}|^2,
\end{displaymath}
where $m_l$ is the mass of the lepton, we can inversely obtain
\begin{eqnarray}
f_{D^+}|V_{cd}| = 46.4 \pm 2.0 \;{\rm MeV}.  \label{Dfvcd}
\end{eqnarray}
Furthermore, using the PDG value for $|V_{cd}| = 0.230 \pm 0.011$~\cite{PDG}, together with Eqs.(\ref{fvcd},\ref{Dfvcd}), we can obtain an experimental constrain for the multiplication of $f_{D^+}$ with $f^{D\to\pi}_+(0)$, i.e.,
\begin{eqnarray}
f_{D^+} f^{D\to\pi}_+(0) = 0.128 \pm 0.012 \;{\rm GeV}.  \label{exffdpi}
\end{eqnarray}
Combining this experimental values (\ref{exffdpi}) of $f_{D^+}f^{D\to\pi}_+(0)$ with the theoretical one calculated by sum rules (\ref{fbpi}), as shown by Fig.(\ref{fig_FFDPI}), we obtain the second reasonable region for the parameter $B$:
\begin{eqnarray}
B_{(D \to \pi l \nu)} = [0.00,0.14].  \label{B_Dpi}
\end{eqnarray}
Fig.(\ref{fig_FFDPI}) gives the value of $f_{D^+}f^{D\to\pi}_+(0)$ versus the parameter $B$, where the shaded band indicates the experimental values (\ref{exffdpi}), the solid, the dashed and the doted lines stand for the central, upper and lower edge of the theoretical values calculated by the LCSR (\ref{fbpi}) with slight parameter changes to agree with the $D$-meson case. Here we have implicitly set the value of $B$ to be bigger than $0$, which is reasonable, since as shown in Fig.(\ref{phi}), by varying $B\in[0,0.6]$ the DA can mimic all of its known behaviors suggested in the literature.

As a final remark, the $D$-meson mass may be not large enough, the energy scale is about $1.5$ GeV, thus, the reliability of the LCSR for the form factor $f^{D\to\pi}_+$ may be less reliable than the $B$-meson case. So we give two schemes for setting the region of parameter $B$:
\begin{itemize}
\item Scheme A: If we believe the LCSR has the same importance as that of $B\to\pi l \nu$, then the range of $B$ is
    \begin{eqnarray}
    B = [0.01,0.14].       \label{B_BDpi}
    \end{eqnarray}
\item Scheme B: If only the LCSR for $B\to\pi l \nu$ is acceptable, we have a broader region as shown in Eq.(\ref{B_Bpi}).
\end{itemize}

\section{Discussion}

\begin{figure}[htb]
\centering
\includegraphics[width=0.48\textwidth]{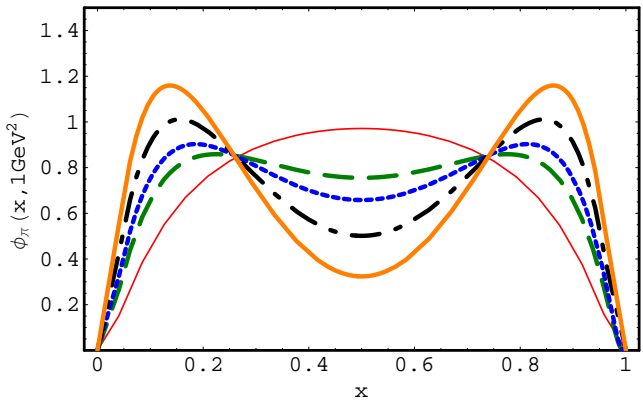}
\caption{The pion leading-twist DA versus the parameter $B$ at $\mu = 1$ GeV, where the thin-solid, the dashed, the dotted, the dash-dotted and the thick-solid lines are $B = 0.01$, $0.14$, $0.2$, $0.3$ and $0.42$, respectively. } \label{DAs}
\end{figure}

If the parameter $B$ is determined, the shape of the pion leading-twist DA can be fixed. Under the scheme A, the second and fourth moments of the pion twist-2 DA can be calculated as $a_2(1 {\rm GeV}) =[0.039,0.184]$ and $a_4(1 {\rm GeV}) =[-0.027,-0.028]$. Under the scheme B, the first two moments changes to $a_2(1 {\rm GeV}) =[0.039,0.495]$ and $a_4(1 {\rm GeV}) =[-0.027,-0.004]$. We present our DA model with different values of $B$ at $\mu = 1$ GeV in Fig.(\ref{DAs}), where the thin-solid line, the dashed line, the dotted line, the dash-dotted line and the thick-solid line are for $B = 0.01$, $0.14$, $0.2$, $0.3$ and $0.42$, respectively.

As two applications, we apply our pion leading-twist DA to deal with the pion-photon TFF $F_{\pi\gamma}(Q^2)$ and the branching ratio of the $B$-meson exclusive decay $B^0 \to \pi^0\pi^0$.

\subsection{The pion-photon TFF $F_{\pi\gamma}(Q^2)$}

As a first application, we revisit the pion-photon TFF. The pion-photon TFF provides the simplest example for the perturbative analysis to exclusive process, which has aroused people's great interest since it was first analyzed by Lepage and Brodsky~\cite{TFFas}. Later on, to explain the abnormal large $Q^2$ behavior observed by the BABAR Collaboration in 2009~\cite{TFF_BABAR}, many works have been done, e.g. by the perturbative QCD approach~\cite{TFF2010,TFF2011,TFF2010_Q0,TFF2011_MPA} or by the LCSR approach~\cite{TFF2009_LCSR,TFF2011_LCSR1,TFF2011_LCSR2}. However, last year, the Belle Collaboration released their new analysis~\cite{TFF_BELLE}, which  dramatically different from those reported by BABAR Collaboration, but likely to agree with the asymptotic behavior estimated by Ref.\cite{TFFas}. Many attempts have been tried to clarify the situation~\cite{DA_fbpi,TFF2012,TFF2012_LCSR1,TFF2012_LCSR2,TFF2012_Q0,TFF2013_LCSR}.

Following the idea suggested by Ref.\cite{PPTFF96}, we have studied the pion-photon TFF with the pQCD approach by carefully dealing with the transverse momentum corrrections~\cite{TFF2007,TFF2010,TFF2011,TFF2012}. Generally, the pion-photon TFF $F_{\pi\gamma}(Q^2)$ can be written as a sum of
the valence quart part $F_{\pi\gamma}^{(V)}(Q^2)$ and the non-valence quark part $F_{\pi\gamma}^{NV} (Q^2)$:
\begin{eqnarray}
F_{\pi\gamma}(Q^2) = F_{\pi\gamma}^{(V)}(Q^2) + F_{\pi\gamma}^{(NV)}(Q^2) .
\label{TFF_total}
\end{eqnarray}
The valence quark part $F^{(V)}_{\pi\gamma}(Q^2)$ indicates the pQCD calculable leading Fock-state contribution, e.g., the direct annihilation of the valence $q\bar{q}$ pair into two photons, which dominates the TFF when $Q^2$ is large. The non-valence quark part $F^{(NV)}_{\pi\gamma}(Q^2)$ is related to the non-perturbative higher Fork-states contributions, which can be estimated via a proper phenomenological model. The analytic expressions for $F_{\pi\gamma}^{(V)}(Q^2)$ and $F_{\pi\gamma}^{NV} (Q^2)$ can be found in Ref.\cite{TFF2010}.

\begin{figure}[tb]
\centering
\includegraphics[width=0.48\textwidth]{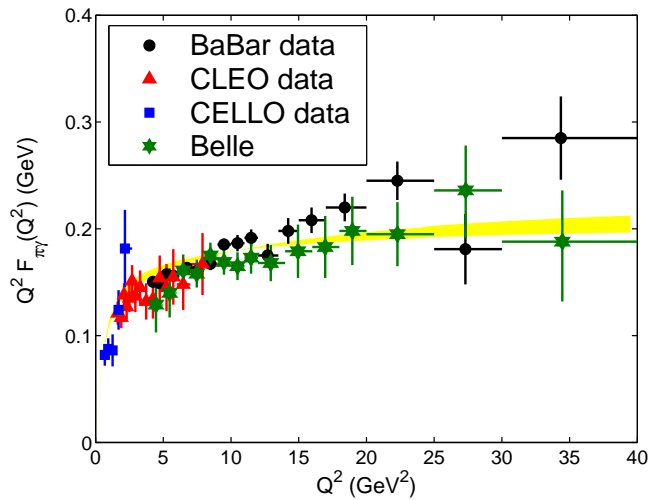}
\caption{$Q^2 F_{\pi\gamma}(Q^2)$ with our WF model (\ref{WF_full},\ref{WF_spatial}) by varying the model parameter $B$ within the region $[0.01,0.14]$. The shaded band is our theoretical estimation.} \label{fig_PPTFF}
\end{figure}

Taking all of the input parameters to be same as those in Ref.\cite{TFF2010}, but with our present DA model with $B\in[0.01,0.14]$, we draw the pion-photon TFF $F_{\pi\gamma}(Q^2)$ in Fig.(\ref{fig_PPTFF}). The upper and lower borderlines correspond to $B=0.01$ and $0.14$ respectively. It shows that in the small $Q^2$ region, $Q^2\lesssim 15~GeV^2$, the pion-photon TFF can explain the CELLO~\cite{TFF_CELLO}, CLEO~\cite{TFF_CLEO}, BABAR~\cite{TFF_BABAR} and Belle~\cite{TFF_BELLE} experimental data simultaneously. While for the large $Q^2$ region, our present estimation favors the Belle data and disfavors the BABAR data. This result is in agreement with the conclusion of Refs.\cite{TFF2012_LCSR2,TFF2013_LCSR}. If taking $B\in[0.01,0.42]$, the calculated curve for the pion-photon TFF with the upper limit of the parameter ($B=0.42$) will be between the Belle and BABAR data.

\subsection{The $B$-meson exclusive decay $B^0 \to \pi^0\pi^0$}

As a second application, we discuss with the process $B^0\to\pi^0\pi^0$, which has been calculated within the pQCD approach~\cite{BPP_2003,BPP_2005,BPP_Li95,BPP01,BPP_Lu01,BPP04}. At present, we adopt the same calculation technology as described in Refs.\cite{BPP_Li95,BPP01,BPP_Lu01,BPP04} to do our calculation. The corresponding decay width can be written as:
\begin{eqnarray}
\Gamma (B^0 \to \pi^0\pi^0) &=& \frac{C_F^2 M_B^2}{128\pi} \left| V^\ast_{ub} V_{ud} M^T_a \right|^2 \nonumber\\
&&\times \left[ 1+z^2+2z \cos(\delta - \gamma) \right], \label{bpipif}
\end{eqnarray}
where $V_{ub} \simeq \left| V_{ub} \right| e^{-i\gamma}$, $z = \left| V^\ast_{tb} V_{td} / V^\ast_{ub} V_{ud} \right| \left| M^P_a / M^T_a \right|$, $\delta$ is the relative strong phase between tree diagrams ($M^T_a$) and penguin diagrams ($M^P_a$), $\gamma$ is the CKM phase angle. The specific corresponding formulas can be found in Ref.\cite{BPP04}.

\begin{figure}[t]
\centering
\includegraphics[width=0.48\textwidth]{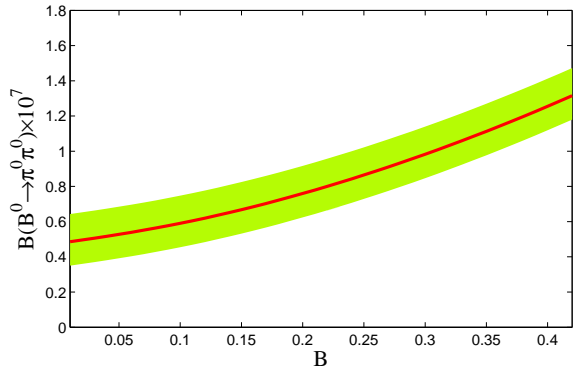}
\caption{The branching ratio $\mathcal{B}(B^0\to\pi^0\pi^0)$ for the exclusive process $B^0\to\pi^0\pi^0$ versus the parameter $B$. The red solid line indicates the central value of $\mathcal{B}(B^0\to\pi^0\pi^0)$, and the shaded band stands for the uncertainty from the dominant uncertainty sources such as $m_0$, $\gamma$, $|V_{ub}|$ and $f_B$. }  \label{fig_ABPIPI}
\end{figure}

In doing the numerical calculation, we adopt the same $B$-meson DA and pion twist-3 DAs used in Refs.\cite{BPP04}, but with our present pion leading-twist DA. The result is shown in Fig.(\ref{fig_ABPIPI}), where we vary the parameter $B$ within the region of $[0.01,0.42]$, and the shaded band indicates the uncertainty from the dominant uncertainty sources such as $m_0=1.6\pm 0.2 \rm GeV$~\cite{m0}, $\gamma ={68^{\circ}}^{+10^{\circ}}_{-11^{\circ}}$~\cite{PDG}, $|V_{ub}|$ and $f_B$. Other parameters are taken as their central values listed in Particle Data Group~\cite{PDG}, e.g. $|V_{td}|=0.00867$, $|V_{tb}|=0.999146$, $|V_{ud}|=0.97427$, $m_{B^0}=5279.58 \rm MeV$, $m_{\pi^0}=134.9766 \rm MeV$, due to their uncertainties are comparatively much small. Moreover, we take $\Lambda^{f=4}_{QCD}=0.25 \rm GeV$ as in Ref.\cite{BPP04}. The branching ratio $\mathcal{B}(B^0\to\pi^0\pi^0)$ increases with increment of the parameter $B$, i.e. the value of $\mathcal{B}(B^0\to\pi^0\pi^0)$ is increasing and is closing to the experimental data for a larger $B$. This agrees with the behavior of the pion leading-twist DA shown in Fig.(\ref{DAs}). Fig.(\ref{DAs}) shows that the pion leading-twist DA in the region closing to the endpoint becomes larger when the parameter $B$ is bigger, and correspondingly the obtained branching ratio $\mathcal{B}(B^0\to\pi^0\pi^0)$ becomes larger. This situation do not imply that there is endpoint singularity for our modal DA. For the twist-three contributions, because of the inclusion of the $k_T$-dependent terms~\cite{BPP_Li95,BPP_Li_other}, our calculation also has no endpoint singularity.

Our present estimation, $\mathcal{B}(B^0\to\pi^0\pi^0) \sim [0.35,1.47]\times10^{-7}$, is much smaller than the experimental data $(1.62 \pm 0.31) \times 10^{-6}$~\cite{PDG}, the reason lies in that I) We only take the LO contribution into consideration. At present, we mostly care about the influence from the twist-2 DA model parameter $B$, and do not expect to solve the puzzle that there is tremendous difference between the experimental data and the theoretical estimation; II) As indicated by Refs.\cite{bpipi1,bpipi2,bpipi3,bpipi4}, there may have some important factors need to be considered in the calculation, such as the next-to-order correction may be big or there may have large non-perturbative contributions, even unknown mechanism may exist, which is beyond the scope of the present paper.

\section{Summary}

In the present paper, based on the revised LC harmonic oscillator model for the pion leading-twist DA, we have made a combined analysis of the pion DA by using the channels $\pi^0 \to \gamma \gamma$, $\pi \to \mu \bar{\nu}$, the semi-leptonic decays $B\to \pi l \nu$ and $D \to \pi l \nu$ in comparison with the experimental data. Based on the constraints from these processes, typical parameters for the pion leading-twist DA are presented in Table \ref{tab2}.

In addition to the two constraints (\ref{DA_constraint1},\ref{DA_constraint2}), by using the constraint from the process $B \to \pi l \nu$, the parameter $B$ is restricted in $[0.01,0.42]$. If taking the process $D \to \pi l \nu$ as a further constrain, we can obtain a more narrow region $B = [0.00,0.14]$. Using the pion leading-twist DA model, we recalculate the branching ratio $\mathcal{B}(B^0\to\pi^0\pi^0)$ and the pion-photon TFF. The branching ratio $\mathcal{B}(B^0\to\pi^0\pi^0)$ increases with increment of the parameter $B$. For the pion-photon TFF, our present result with the parameter $B = [0.01,0.14]$ favors the Belle data and the corresponding pion DA has the slight difference from the asymptotic form. Then, one can predict the behavior of the pion-photon TFF in high $Q^2$ regions which can be tested in the future experiments. It is expected that BABAR and Belle can obtain more accurate and consistent data in the future, then the behavior of the pion DA can be further determined completely. On the other hand, we can adopt more pionic processes, such as the pion electromagnetic form factor, to make a further constrain to the pion DA, which is in progress. It is believed that the pion DA will be determined by the global fit to the exclusive processes involving the pion in the coming future.

\vspace{0.5cm}

{\bf Acknowledgments}: The authors would like to thank Zuo-Hong Li, Nan Zhu and Zhi-Tian Zou for helpful discussions. This work was supported in part by Natural Science Foundation of China under Grant No.11235005 and No.11075225, and by the Program for New Century Excellent Talents in University under Grant NO.NCET-10-0882.

\end{document}